\documentclass[aps,preprint,prd,eqsecnum,showpacs,letterpaper]{revtex4}
\usepackage{graphicx}
\usepackage{amsmath}
\usepackage{color}
\usepackage{subfig}
\usepackage{multirow}
\usepackage{slashed}
\usepackage{hyperref}

\makeatletter
\newbox\slashbox \setbox\slashbox=\hbox{$/$}
\newbox\Slashbox \setbox\Slashbox=\hbox{\large$/$}
\def\pFMslash#1{\setbox\@tempboxa=\hbox{$#1$}
  \@tempdima=0.5\wd\slashbox \advance\@tempdima 0.5\wd\@tempboxa
  \copy\slashbox \kern-\@tempdima \box\@tempboxa}
\def\pFMSlash#1{\setbox\@tempboxa=\hbox{$#1$}
  \@tempdima=0.5\wd\Slashbox \advance\@tempdima 0.5\wd\@tempboxa
  \copy\Slashbox \kern-\@tempdima \box\@tempboxa}

\def\miss#1{\ifmmode{/\mkern-11mu #1}\else{${/\mkern-11mu #1}$}\fi}
\makeatother

\def\dal{\,\lower0.9pt\vbox{\hrule \hbox{\vrule height 0.35 cm \hskip 0.3 cm
\vrule height 0.35 cm}\hrule}\,}

\begin{document}

\title{Flavor violation in chromo- and electromagnetic dipole moments induced by $Z^\prime$ gauge bosons and a brief revisit of the Standard Model}

\author{J. I. Aranda$^a$, D. Espinosa-G\'omez$^a$, J. Monta\~no$^{a,b}$, B. Quezadas-Vivian$^a$, F. Ram\'irez-Zavaleta$^a$, and E. S. Tututi$^a$}

\affiliation{$^a$Facultad de Ciencias F\'{\i}sico Matem\'aticas, Universidad Michoacana de San Nicol\'as de Hidalgo,
Avenida Francisco J. M\'{u}jica S/N, 58060,  Morelia, Michoac\'an, M\'exico\\
$^b$CONACYT, M\'exico}

\begin{abstract}
The electromagnetic dipole moments of the tau lepton and the chromoelectromagnetic dipole moments of the top quark are estimated via flavor-changing neutral currents, mediated by a new neutral massive gauge boson. We predict them in the context of models beyond the Standard Model with extended current sectors, in which simple analytic expressions for the dipole moments are presented. For the different $Z^\prime$ gauge boson considered, the best prediction for the magnetic dipole moment of the tau lepton, $|a_\tau|$, is of the order of $10^{-8}$, while the highest value for the electric one, $|d_\tau|$, corresponds to $10^{-24}$ $e\,$cm; our main result for the chromomagnetic dipole moment of the top quark, $|\hat{\mu}_t|$, is $10^{-6}$, and the value for the chromoelectric one, $|d_t|$, can be as high as $10^{-22}$ $e\,$cm. We compare our results, revisiting the corresponding Standard Model predictions, in which the chromomagnetic dipole moment of the top quark is carefully evaluated, finding explicit imaginary contributions.
\end{abstract}

\vskip2.5pc
\maketitle

\section{Introduction}
In the Standard Model (SM), flavor-changing transitions promoted by neutral gauge bosons can be found in the quark sector; however, these are strongly suppressed by the GIM mechanism and because they are induced at the one-loop level~\cite{SMqsectorsup}. On the other hand, in the leptonic sector Lagrangian, the SM contains an exact flavor symmetry, which implies that transitions between charged leptons mediated by neutral gauge bosons are forbidden to any perturbative order. Although in the SM the flavor violation phenomenon is suppressed, it is known that the impact of flavor-changing neutral currents (FCNCs) could be increased by new physics effects due, for example, to both extended Yukawa~\cite{EYS} or new current sectors~\cite{arhrib,NCS1,NCS2}. The study of flavor violation has gained much interest due to the discovery of neutrino oscillations~\cite{neutrinos}. However, this phenomenon occurs exclusively between neutral fermions (neutrinos), and therefore transitions between charged leptons would play a complementary role by offering clear signals of flavor violation, enriching such a phenomenon. According to this, the proposal of this work is to study the effects of new physics on the electromagnetic and chromoelectromagnetic properties of charged fermions due to the presence of FCNCs mediated by a new neutral massive gauge boson identified as $Z^\prime$. The existence of this boson has been proposed in numerous extended models, the simplest those being ones that involve an extra $U^\prime(1)$ gauge symmetry group~\cite{langacker1}. The simplest model that predicts the existence of the $Z^\prime$ boson is founded on the $SU_L(2)\times U_Y(1)\times U^\prime(1)$ extended electroweak gauge group~\cite{robinett,langacker2,leike,perez-soriano}.

At present, the experimental collaborations ATLAS and CMS, at the LHC, have devoted many studies to the search for new elementary particles, such as new neutral massive gauge bosons~\cite{ATLAS1,CMS1} or new scalar bosons~\cite{Scalars-ATLAS-CMS}. As far as the search for new neutral massive gauge bosons is concerned, the experimental results indicate that the existence of $Z^\prime$ bosons is not excluded for masses slightly above 3 TeV. Specifically, the ATLAS Collaboration establishes lower limits on the $Z^\prime$ masses ranging from 2.74 up to 3.36 TeV at 95\% C.L.~\cite{ATLAS1,PDG}. In contrast, the CMS Collaboration reports that the existence of $Z^\prime$ gauge bosons would be excluded for masses below the range between 2.57 and 2.9 TeV at 95\% C.L.~\cite{CMS1,PDG}.

The flavor violation (FV) issue has allowed us to relate the hypothetical $Z^\prime$ particle with several processes such as single top production~\cite{arhrib,NCS1}, the $D^0-\bar{D^0}$ mixing system~\cite{NCS1}, the $b^0_q-\bar{b^0_q}$ mixing system~\cite{sahoo1}, lepton flavor-violating decays~\cite{perez-soriano,NCS2,cabarcas,yue}, etc. In this way, by using the most general renormalizable Lagrangian that includes FV mediated by a new neutral massive gauge boson, we will estimate the impact of FCNC on the electromagnetic dipole moments of the tau lepton and the chromoelectromagnetic dipole moments of the top quark, resorting to different grand unification models (GUT) with extended current sectors~\cite{langacker-rmp, arhrib}.

The static magnetic properties of charged leptons in the context of the SM have developed the predictive power of this theory~\cite{amdmtaureview}. However, little is known about the static electric properties of charged leptons. The experimental measurement of the magnetic dipole moment (MDM) of the electron ($a_e$) has been the main argument to establish the SM as a rather successful theory. In contrast, although the MDM of the muon ($a_\mu$) has been studied exhaustively, a discrepancy persists between the experimental measurement~\cite{g-2-e} and the SM theoretical prediction~\cite{g-2-t}, which turns out to be around three standard deviations~\cite{g-2-d}. Therefore, new measurements will be carried out in order to increase the experimental precision and look for possible systematic errors~\cite{g-2-ne}. At the same time, theoretical efforts are realized in order to try to reduce the uncertainty in the theoretical prediction coming from hadronic light-by-light contributions~\cite{g-2-d,g-2-nt}. If such a discrepancy were reduced, it would imply that possible new physics effects would be very restricted. On the other hand, there is practically no information regarding the static electromagnetic properties of the tau lepton, mainly due to its short lifetime~\cite{amdmtaureview}. For the tau magnetic dipole moment there are only experimental bounds, that restrict it with enormous uncertainty, $-0.052<a_\tau<0.013$ at 95\% C.L.~\cite{PDG}. In this sense, we have revisited the so-called SM electroweak contribution for the tau lepton MDM. Similarly, given that for the electric dipole moment (EDM) of charged leptons there are only experimental bounds on their real value, we turn our attention to the EDM of the tau lepton as a source of study of possible new physics effects, related to FV, and given its nature, it would also be related to $\emph{CP}$ violation. Since the SM does not predict appreciable effects of $\emph{CP}$ violation in the leptonic sector~\cite{Belle}, the study of the tau EDM is an ideal testing ground for the search of new physics effects. The experimental measurement attempts of the tau EDM have resulted in the following constraints~\cite{PDG,Belle}: $-2.2\times10^{-17}\;\; e\, \mathrm{cm}<\mathrm{Re}(d_\tau)<4.5\times10^{-17}\;\; e\,\mathrm{cm}$ and $-2.5\times10^{-17}\;\; e\, \mathrm{cm}<\mathrm{Im}(d_\tau)<8.0\times10^{-19}\;\; e\,\mathrm{cm}$. Studies on the EDM have been carried out in Refs.~\cite{GutierrezRodriguez:2006hb,GutierrezRodriguez:2009ns,Gutierrez-Rodriguez:2013eaa}.

Moreover, given the great mass of the top quark, 173 GeV~\cite{PDG}, which is of the order of the Fermi scale, it is thought that this particle could be related to new physics effects present at the TeV energy scale. Thereby, it is interesting to study the physical properties of this particle, our proposal being the characterization of possible flavor-violating effects due to the presence of FCNCs, which would be impacting the chromoelectromagnetic properties of the top quark. Because in the SM the chromomagnetic dipole moment (CMDM) of the top quark appears at the one-loop level and its chromoelectric dipole moment (CEDM) arises at three-loop level, the impact of new physics effects becomes relevant. In addition, appreciable new physics effects on the top CEDM are of great importance as they would directly impact the $\emph{CP}$ violation phenomenon, which would be indicative of new sources of $\emph{CP}$ violation and, in our case, of FV. Currently, the spin correlations of top-antitop pairs and the polarization of the top quark have been measured in pp collisions at $\sqrt{s}=8$ TeV~\cite{Bounds-CMDM-CEDM}. These results were obtained by the CMS Collaboration at CERN, where constraints on extended models are imposed, finding new exclusion limits at 95\% of C.L. for the CMDM and CEDM of the top quark, namely, $-0.053<\mathrm{Re}(\hat{\mu}_t)<0.026 $ and $-0.068<\mathrm{Im}(\hat{d}_t)<0.067$~\cite{Bounds-CMDM-CEDM}, respectively. The top-quark CMDM and CEDM have been calculated in the SM \cite{Atwood:1994vm}, as well as in other extensions such as the two-Higgs doublet model \cite{Gaitan:2015aia}, the minimal supersymmetric Standard Model \cite{Martinez:2001qs,Aboubrahim:2015zpa}, 3-3-1 models \cite{Martinez:2007qf}, technicolor models \cite{Appelquist:2004es}, models with vectorlike multiplets \cite{Ibrahim:2011im}, effective operators \cite{Hayreter:2013kba}, and the two-Higgs doublet model with four fermion generations~\cite{Tavares}. However, the SM CMDM contribution of the top quark coming from the three-gluon vertex is in fact divergent when the gluon is on shell, but in Ref.~\cite{Martinez:2007qf} the authors claim that it is finite. Indeed, Refs.~\cite{Choudhury:2014lna} and \cite{Bermudez:2017bpx} are in agreement with the ill behavior when the gluon is on shell. In view of such an issue we were forced to revisit in depth the complete one-loop SM calculations for the CMDM of the top quark, finding novelties that will be commented on below.

The rest of this paper is organized as follows. In Sec.~\ref{sec2}, the basis of FCNCs induced by a new neutral massive gauge boson of spin 1 is presented, where it is explained how bounds over $Z^\prime f_if_j$ (for $f_if_j=\tau\mu,\tau e, tc, tu$) couplings are determined. In Sec. III, we exhibit the theoretical results for the electromagnetic and chromoelectromagnetic dipole moments induced by FCNCs. Also, we present the numerical analysis for the MDM (CMDM) and the EDM (CEDM) of the tau lepton (top quark), respectively; in addition, we present a brief revisit of the CMDM of the top quark in the SM. Finally, Sec. IV gives the conclusions.

\section{Theoretical framework}\label{sec2}

Since it is required to estimate the strength of the $Z^\prime f_if_j$ couplings (where $f_ {i, j}$ represents any SM charged fermion) in order to determine its impact on the MDM, EDM, CMDM, and CEDM, it is necessary establish the Lagrangian that comprises FCNCs mediated by the $Z^\prime$ gauge boson. The most general renormalizable Lagrangian that includes FV mediated by a new neutral massive gauge boson, coming from any extended model or GUT~\cite{durkin, langacker3, Salam-Mohapatra}, is
\begin{equation}\label{lnc}
\mathcal{L}_{NC}=\sum_{i,j}\left[\, \overline{f}_i\,
\gamma^{\alpha} (\Omega_{{L}f_if_j} \, P_L+\Omega_{{R}f_if_j} \,
P_R)\, f_j+\overline{f}_j\, \gamma^{\alpha} ({\Omega^\ast_{{L}f_jf_i}}\,
P_L+{\Omega^\ast_{{R}f_jf_i}} \, P_R) \, f_i \, \right]Z^{\prime}_{\alpha},
\end{equation}
where $f_{i}$ is any fermion of the SM, $P_{L,R}=\frac{1}{2}(1\pm \gamma_{5})$ are the chiral projectors, and $Z^\prime_\alpha$ is a new neutral massive gauge boson predicted by several extensions of the SM~\cite{durkin, langacker3, Salam-Mohapatra, Pleitez}. The $\Omega_{Lf_if_j}$, $\Omega_{Rl_il_j}$ parameters represent the strength of the $Z^\prime f_if_j$ coupling, where $f_i$ is any charged fermion of the SM. From now on, we will assume that $\Omega_{Lf_if_j}=\Omega_{Lf_jf_i}$ and $\Omega_{Rf_if_j}=\Omega_{Rf_jf_i}$. The Lagrangian in Eq.~(\ref{lnc}) includes both flavor-conserving and flavor-violating couplings mediated by a $Z^\prime$ gauge boson. In this work, the following $Z^\prime$ bosons are considered: the $Z_S$ of the sequential $Z$ model, the $Z_{LR}$ of the left-right symmetric model, the $Z_\chi$ boson that arises from the breaking of $SO(10)\to SU(5)\times U(1)$, the $Z_\psi$ that emerges as a result of $E_6\to SO(10)\times U(1)$, and the $Z_\eta$ appearing in many superstring-inspired models~\cite{langacker2}. Concerning to the flavor-conserving couplings, $Q^{f_i}_{L,R}$~\cite{robinett,langacker2,arhrib}, the values of these are shown in Table~\ref{table1}, for different extended models are related to the $\Omega$ couplings as $\Omega_{Lf_if_i}=-g_2 \,Q^{f_i}_L$ and $\Omega_{Rf_if_i}=-g_2 \,Q^{f_i}_R$, where $g_2$ is the gauge coupling of the $Z^\prime$ boson. For the extended models we are interested in, the gauge couplings of $Z^\prime$'s are
\begin{equation}
g_2=\sqrt{\frac{5}{3}}\sin\theta_W g_1\lambda_g,
\end{equation}
where $g_1=g/\cos\theta_W$, $\lambda_g$ depends on the symmetry breaking pattern being of $\mathcal{O}(1)$~\cite{robinett2}, and $g$ is the weak coupling constant. In the sequential $Z$ model, the gauge coupling $g_2=g_1$.
\begin{table}[htb!]
\caption{\label{table1}
Chiral-diagonal couplings of the extended models.}
\medskip
\begin{ruledtabular}
\begin{tabular}{cccccc}
      & $Z_S$ & $Z_{LR}$ & $Z_\chi$ & $Z_\psi$ & $Z_\eta$ \\
\hline
$Q_L^{l_i}$ & $-0.2684$ & $0.2548$ &  $\frac{3}{2\sqrt{10}}$ &
      $\frac{1}{\sqrt{24}}$ &  $\frac{1}{2\sqrt{15}}$ \\
$Q_R^{l_i}$ & $0.2316$ & $-0.3339$ &  $\frac{-3}{2\sqrt{10}}$ &
      $\frac{-1}{\sqrt{24}}$ & $\frac{-1}{2\sqrt{15}}$ \\
$Q_L^{u_i}$ & $0.3456$ & $-0.08493$ &  $\frac{-1}{2\sqrt{10}}$ &
      $\frac{1}{\sqrt{24}}$ &  $\frac{-2}{2\sqrt{15}}$ \\
$Q_R^{u_i}$ & $-0.1544$ & $0.5038$ &  $\frac{1}{2\sqrt{10}}$ &
      $\frac{-1}{\sqrt{24}}$ & $\frac{2}{2\sqrt{15}}$ \\
\end{tabular}
\end{ruledtabular}
\end{table}

\subsection{Bounding the $Z^\prime f_if_j$ couplings}
The subject of this work is to study the impact of flavor-violating couplings mediated by a $Z^\prime$ gauge boson on the MDM and the EDM of the tau lepton, and the CMDM and the CEDM of the top quark. To do this task, we will use bounds on the lepton flavor-violating couplings $Z^\prime\tau\mu$ and $Z^\prime\tau e$, which have been previously computed by using the experimental constraints for the lepton flavor-violating $\tau\to\mu\mu^+\mu^-$ and  $\tau\to\mu e^+e^-$ decays~\cite{NCS2}. Finally, we will use the results of a previous work in which the strength of the $Z^\prime tc, Z^\prime tu$ couplings is estimated by means of the $D^0-\bar{D^0}$ mixing system~\cite{NCS1}.

\subsubsection{Three-body $\tau\to\mu\mu^+\mu^-,ee^+e^-$ decays}
\begin{figure}[htb!]
\centering
\includegraphics[width=6cm]{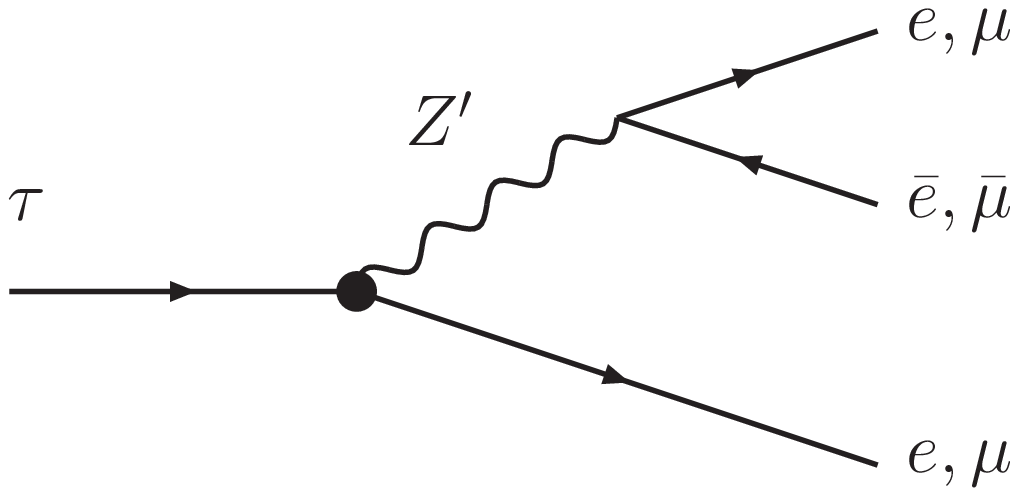}
\caption{\label{tmmm} Feynman diagrams corresponding to the $\tau\to\mu\mu^+\mu^-$ and $\tau\to ee^+e^-$ decays.}
\end{figure}

The contribution of the flavor-violating $Z^\prime l_il_j$ vertex to the $\tau\to l_i{l_i}^+{l_i}^-$ decay is depicted in Fig.~\ref{tmmm}, where $l_il_j$ represents $\tau\mu$ or $\tau e$ and $l_i{l_i}^+{l_i}^-$ symbolizes $\mu \mu^+\mu^-$ or $e e^+e^-$. The three-body decay of the tau lepton comes from the tree-level Feynman diagram, whose associated branching ratio was computed in a previous work~\cite{NCS2}
\begin{align}\label{br3}
Br(\tau\to l_i{l_i}^+{l_i}^-)=\frac{g_2^2}{384 \pi^3}h_1(m_{Z^\prime})(|Q^{e}_{L}\Omega_{L\,l_il_j}|^2+|Q^{e}_{R}\Omega_{R\,l_il_j}|^2)\,\frac{m_\tau}{\Gamma_\tau},
\end{align}
where
\begin{equation}
h_1(m_{Z^\prime})=\int_0^1dx\frac{2x-1}{(x-1+m^2_{Z^\prime}/m_\tau^2)^2}(2(7-4x)x-5),
\end{equation}
and $\Gamma_\tau$ is the total decay width of the tau lepton. The branching ratio in Eq.~(\ref{br3}) must be less than the corresponding experimental bounds to the processes $\tau\to \mu\mu^+\mu^-$ and $\tau\to ee^+e^-$, as applicable. It is considered that $\mathrm{Br_{Exp}}(\tau\to \mu\mu^+\mu^-)<2.1\times10^{-8}$~\cite{PDG} and $\mathrm{Br_{Exp}}(\tau\to ee^+e^-)<2.7\times10^{-8}$~\cite{PDG}, which allow us to get constraints on the flavor-violating  parameters: $|\Omega_{L\tau\mu}|^2, |\Omega_{R\tau\mu}|^2$, $|\Omega_{L\tau e}|^2, |\Omega_{R\tau e}|^2$.

\subsubsection{$D^0-\bar{D^0}$ mixing system}
For FCNCs mediated by a new neutral massive gauge boson, in a previous work~\cite{NCS1} the mass difference, $\Delta M_{D}$, coming from the $D_0-\overline{D_0}$ mixing system, was estimated. Explicitly, $\Delta M_{D}$ can be written as
\begin{align}
\Delta M_D&=\frac{1}{12}\frac{\Omega_{uc}^2}{m^2_{Z^\prime}} f_D^2 M_D B_D \bigg[1+\frac{x}{8\pi^2}(32f(x)-5g(x))\bigg],
\end{align}
where $B_D$ is the bag model parameter and $f_D$ symbolizes the $D_0$-meson constant decay. Here, we are taking $B_D\sim 1$, $f_D=222.6$ MeV~\cite{mar}, and $M_D=1.8646$ GeV~\cite{PDG}. By assuming that $\Delta M_{D}$ does not exceed the experimental uncertainty, we are able to constraint the $\Omega_{uc}$ parameter~\cite{NCS1}
\begin{align}\label{ec2}
  |\Omega_{uc}|<\frac{3.6\times 10^{-7}m_{Z'}\text{GeV}^{-1}}{\sqrt{1+\frac{x}{8\pi^2}(32f(x)-5g(x))}}.
\end{align}
From this bound, we can estimate the $\Omega_{tc}$ and $\Omega_{tu}$ parameters by considering that $|\Omega_{uc}|\approx |\Omega_{tc}\Omega_{tu}|$ and $\Omega_{tc}=10\Omega_{tu}$; the details of the calculation and the justification for such assumptions can be found in Ref.~\cite{NCS1}. Therefore, the coupling parameters are given as
\begin{align}\label{ecD0}
  |\Omega_{tc}|^2 &< \frac{3.6\times 10^{-6}m_{Z'}\text{GeV}^{-1}}{\sqrt{1+\frac{x}{8\pi^2}(32f(x)-5g(x))}},\nonumber\\
  |\Omega_{tu}|^2 &<\frac{3.6\times 10^{-8}m_{Z'}\text{GeV}^{-1}}{\sqrt{1+\frac{x}{8\pi^2}(32f(x)-5g(x))}}.
\end{align}

It is pertinent to comment that another possibility for bounding flavor-violating couplings is that coming from experimental limits on the electric dipole moment of the neutron~\cite{Jordy}.

\section{Results and discussion}
In this section, we exhibit the analytical results for the MDM, EDM, CMDM, and CEDM induced by FCNCs mediated by the $Z^\prime$ gauge boson. Subsequently, the corresponding numerical results will be presented.
\begin{center}
\begin{figure}[!h]
\subfloat[]{\includegraphics[width=5cm]{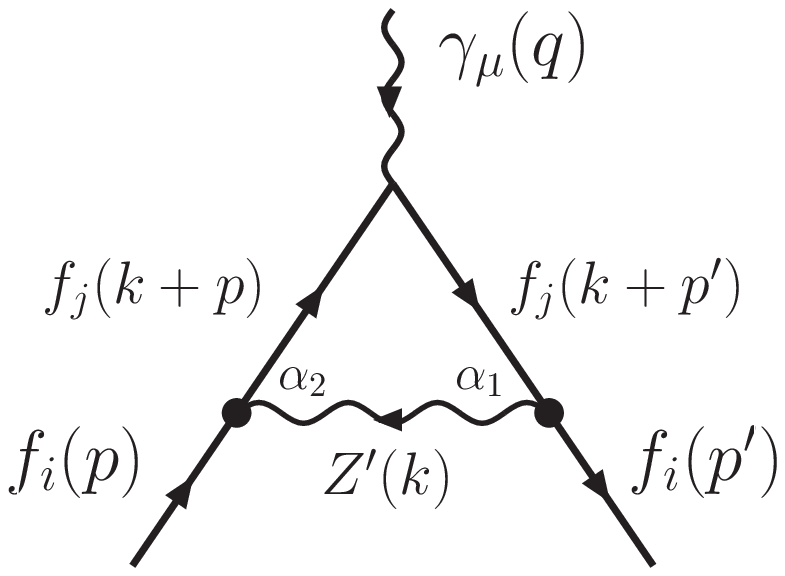}} \qquad
\subfloat[]{\includegraphics[width=5cm]{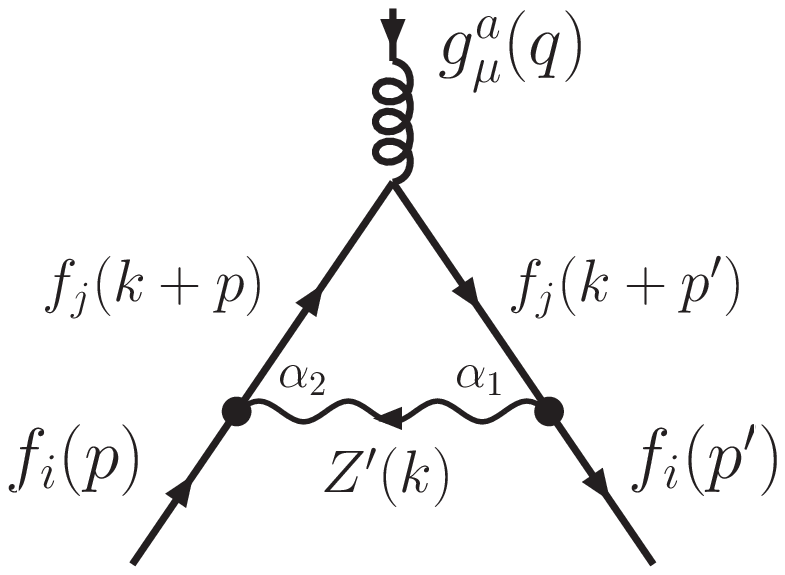}}
\caption{(a) Electromagnetic ($f=l$) and (b) chromoelectromagnetic ($f=q$) dipole moments at one-loop level mediated by a $Z'$ gauge boson with FV.}
\label{FIGURE-1-loop}
\end{figure}
\end{center}

\subsection{Static electromagnetic dipole moments}
The effective electromagnetic dipole moment Lagrangian for charged leptons, $f=l$, is
\begin{align}\label{lagrangian-leptons}
\mathcal{L}_\text{eff}=&\ -\frac{1}{2}\bar{f}\sigma^{\mu\nu}(F_M+iF_E\gamma^5)f F_{\mu\nu}~,
\end{align}
where $F_M$ is the magnetic form factor and $F_E$ is the electric form factor, $\sigma^{\mu\nu}\equiv\frac{i}{2}[\gamma^\mu,\gamma^\nu]$, and $F_{\mu\nu}=\partial_\mu A_\nu-\partial_\nu A_\mu$ is the photon field strength. The associated vertex is
\begin{equation}\label{}
\Gamma^\mu=\sigma^{\mu\nu}q_\nu \left(F_M+iF_E\gamma^5\right).
\end{equation}
On the other hand, the invariant amplitude is
\begin{equation}\label{}
\mathcal{M}=\mathcal{M}^\mu\epsilon_\mu(\vec{q})~,
\end{equation}
being $\mathcal{M}^\mu=\bar{u}(p')\Gamma^\mu u(p)$.

The static properties arise when the photon is on shell, $q^2=0$, and hence the static anomalous magnetic, $a_f$, and electric, $d_f$, dipole moments \cite{Roberts:2010zz} are
\begin{equation}\label{electromagnetic-lepton}
F_M\equiv \frac{eQ_{f}}{2m_{f}} a_{f}
\quad , \quad
F_E\equiv Q_{f}d_{f} \ .
\end{equation}
It is usual to express them as a single complex dipole form factor,
\begin{equation}\label{}
F_C=F_M+iF_E=|F_C|e^{i\phi_f} ,
\end{equation}
with
\begin{equation}\label{chiral-phase}
|F_C|=\sqrt{F_M^2+F_E^2}
\quad , \quad
\tan\phi_f=\frac{F_E}{F_M}~ ,
\end{equation}
where $\phi_{f}$ is the phase that parametrizes the relative size of the EDM and its MDM.

To compare the results derived in this section we have also calculated the corresponding SM contributions at one-loop level to the tau MDM. Our approximate analytical expressions, which excellently agree with the complete calculations, are
\begin{align}\label{magnetic-dipoles-SM}
a_{l_i}(\gamma)=&~\frac{\alpha}{2\pi} \ ,
\\
a_{l_i}(W)\simeq&~\frac{5G_Fm_{l_i}^2}{12\sqrt{2}\pi^2} \ ,
\\
a_{l_i}(Z)\simeq&~
\frac{G_Fm_{l_i}^2}{6\sqrt{2}\pi^2} \frac{(1-4s_W^2)^2-5}{4} \ ,
\\
a_{l_i}(H)\simeq&~-\frac{G_Fm_{l_i}^2}{24\sqrt{2}\pi^2}\frac{m_{l_i}^2}{m_H^2}
\left(7+6\log{\frac{m_{l_i}^2}{m_H^2}}\right) .
\end{align}
These are valid for any charged lepton and can be compared with those given for the muon
in Sec.~4.2.1 of Ref.~\cite{Jegerlehner:2017gek}. Notice that in our expression for the Higgs contribution we also conserve the first term, which is not relevant for the electron and muon cases but it is important for the tau lepton.
The numerical values are given in Table~\ref{TABLE-magnetic-dipoles-SM}, where
the electroweak contribution means $a_{l_i}(\text{EW})=a_{l_i}(W)+a_{l_i}(Z)+a_{l_i}(H)$.

\begin{table}[!h]
  \centering
\begin{tabular}{|c|c|c|c|}\hline
Contribution &  $a_\tau$ \\
\hline
$\gamma$     & $1.16\times10^{-3}$  \\
$W$          & $1.10\times10^{-6}$  \\
$Z$          & $-5.48\times10^{-7}$ \\
$H$          & $9.76\times10^{-10}$ \\
EW           & $5.52\times10^{-7}$  \\
\hline
\end{tabular}
\caption{Anomalous magnetic dipole moment of the tau lepton at one loop in the SM with $m_H=125.18$ GeV \cite{PDG}.}
\label{TABLE-magnetic-dipoles-SM}
\end{table}

\subsection{One-loop $Z'$ contribution to the static electromagnetic and chromoelectromagnetic dipole moments}
\label{SECTION-loop-diagram}
In analogy to the SM $\bar{f}fZ$ coupling, for the $\bar{f}_if_jZ'$ coupling we rewrite this as
\begin{equation}\label{}
\Omega_{Lf_if_j}P_L+\Omega_{Rf_if_j}P_R=g_{VZ'}^{f_if_j}-g_{AZ'}^{f_if_j}\gamma^5 .
\end{equation}
\begin{equation}\label{}
g_{VZ'}^{f_if_j}\equiv \frac{1}{2}(\Omega_{Lf_if_j}+\Omega_{Rf_if_j})
\quad , \quad
g_{AZ'}^{f_if_j}\equiv \frac{1}{2}(\Omega_{Lf_if_j}-\Omega_{Rf_if_j})\gamma^5.
\end{equation}

The general one-loop quantum fluctuation that generates the static electromagnetic dipole moments, depicted in Fig.~\ref{FIGURE-1-loop}, is
\begin{align}\label{loop}
\mathcal{M}_{f_if_j}^\mu&=eQ_{f_j}\int\frac{d^4k}{(2\pi^4)}
\frac{\bar{u}(p')\gamma^{\alpha_1}(g_{VZ'}^{f_if_j}-g_{AZ'}^{f_if_j}\gamma^5)
(\slashed{k}+\slashed{p'}+m_{f_j})\gamma^{\mu}(\slashed{k}+\slashed{p}+m_{f_j})}{(k^2-m_{Z'}^2)[(k+p')^2-m_{f_j}^2][(k+p)^2-m_{f_j}^2]}
\nonumber\\
&\times\gamma^{\alpha_2}(g_{VZ'}^{f_if_j*}-g_{AZ'}^{f_if_j*}\gamma^5)
u(p)\left(-g_{\alpha_1\alpha_2}+\frac{k_{\alpha_1}k_{\alpha_2}}{m_{Z'}^2}\right) .
\end{align}
For the chromoelectromagnetic case, the factor $eQ_{f_j}$ must be replaced by $g_sT^a$. From this loop integral, the complete analytical results for the static electromagnetic dipole moments can be obtained, given in terms of the form factors $F_{M,E}(q^2=0)$; nevertheless, we present more suitable approximate expressions that have been cross-checked, matching excellently.

The MDM form factor is
\begin{align}\label{magnetic}
F_{Mf_if_j}&\simeq
\frac{eQ_{f_j}}{48 \pi^2 m_{Z'}^4}
\Bigg\{
|g_{VZ'}^{f_if_j}|^2\left[m_{f_i}(3  m_{f_j}^2-4m_{Z'}^2)+6 m_{f_j} m_{Z'}^2\right]\nonumber\\
&+|g_{AZ'}^{f_if_j}|^2\left[m_{f_i}(3  m_{f_j}^2-4m_{Z'}^2)-6 m_{f_j} m_{Z'}^2\right]
\Bigg\} ,
\end{align}
where
\begin{align}\label{magnetic-couplings}
|g_{VZ'}^{f_if_j}|^2=&~
\frac{1}{4} \left[(\text{Re}\Omega_{Lf_if_j}+\text{Re}\Omega_{Rf_if_j})^2
+(\text{Im}\Omega_{Lf_if_j}+\text{Im}\Omega_{Rf_if_j})^2\right] ,
\nonumber\\
|g_{AZ'}^{f_if_j}|^2=&~
\frac{1}{4} \left[(\text{Re}\Omega_{Lf_if_j}-\text{Re}\Omega_{Rf_if_j})^2
+(\text{Im}\Omega_{Lf_if_j}-\text{Im}\Omega_{Rf_if_j})^2\right] .
\end{align}
Correspondingly, the EDM form factor is
\begin{equation}\label{electric}
F_{Ef_if_j}\simeq
\frac{ieQ_{f_j}m_{f_j}}{8\pi^2m_{Z'}^2}(g_{VZ'}^{f_if_j}g_{AZ'}^{f_if_j*}-g_{AZ'}^{f_if_j}g_{VZ'}^{f_if_j*})~,
\end{equation}
where
\begin{align}\label{electric-couplings}
g_{VZ'}^{f_if_j}g_{AZ'}^{f_if_j*}-g_{AZ'}^{f_if_j}g_{VZ'}^{f_if_j*}=&~
i (\text{Re}\Omega_{Lf_if_j}\text{Im}\Omega_{Rf_if_j}
-\text{Re}\Omega_{Rf_if_j}\text{Im}\Omega_{Lf_if_j}).
\end{align}

\subsubsection{$CP$ property}
The electromagnetic dipole moments can be distinguished in two scenarios due to the $\emph{CP}$ property:

i) The \emph{CP} conservation (\emph{CP}-c) case, which only allows $a_{fi}$ ($d_{f_i}$ is forbidden), can happen when
\begin{equation}\label{CPC-scenario}
\text{Re}\Omega_L\neq 0, \ \text{Im}\Omega_L\neq 0,\ \text{Re}\Omega_R=0,\ \text{Im}\Omega_R=0.\nonumber
\end{equation}

ii) The \emph{CP} violation (\emph{CP}-v) case, that gives rise to both $a_{fi}$ and $d_{f_i}$, can occur when
\begin{equation}\label{CPV-scenario}
\text{Re}\Omega_L\neq 0, \ \text{Im}\Omega_L= 0,\ \text{Re}\Omega_R=0,\ \text{Im}\Omega_R\neq 0.\nonumber
\end{equation}

\subsection{Predictions on the tau electromagnetic dipole moments}
In this section, we carry out the phenomenological analysis on the tau MDM and EDM by considering the different $Z'$ gauge bosons, $Z'_S$, $Z'_{LR}$, $Z'_\chi$, $Z'_\psi$, and $Z'_\eta$, whose coupling parameters, $\Omega_{L,R}$, were computed in Ref.~\cite{NCS2}.

The tau MDM is conformed by
\begin{equation}\label{magnetic-tau}
a_\tau=a_{\tau e}+a_{\tau\mu}+a_{\tau\tau}~,
\end{equation}
where $a_{l_il_j}$ are given in Eq.~(\ref{electromagnetic-lepton}) in terms of $F_{Mf_if_j}$, the explicit expression of which were given in Eq.~(\ref{magnetic}).

Otherwise, the tau EDM contributions are
\begin{equation}\label{electric-tau}
d_\tau=d_{\tau e}+d_{\tau\mu}+d_{\tau\tau}~,
\end{equation}
where $d_{l_il_j}$ are given in Eq.~(\ref{electromagnetic-lepton}). The explicit expressions for the $F_{Ef_if_j}$ form factors are given in Eq.~(\ref{electric}). Below we are going to analyze the EDM in $e$cm units, as it is common in the literature.

\subsubsection{CP conservation: $a_\tau$}
For the \emph{CP}-c analysis we follow the scenario: $\text{Re}\Omega_L\neq 0, \ \text{Im}\Omega_L\neq 0,\ \text{Re}\Omega_R=0,\ \text{Im}\Omega_R=0$. Here, $a_\tau$ is provided by Eq.~(\ref{magnetic-tau}); the $a_{\tau e}$ and $a_{\tau\mu}$ quantities receive contributions from the coupling parameters, $\Omega_{L, R\tau e}$ and $\Omega_{L, R\tau\mu}$, which can be derived from Eq.~(\ref{br3}) (for more details, see Ref.~\cite{NCS2}), and $a_{\tau\tau}$ depends on the $\Omega_{L, R\tau\tau}$ parameter~\cite{NCS2}. Regarding the $Z'$ boson mass we are going to explore the mass interval, $m_{Z'}=[2.5,5]$ TeV, which respects the current experimental bounds on the $Z^\prime$ boson mass~\cite{PDG}.
\begin{figure}[htb!]
\centering
\includegraphics[width=17cm]{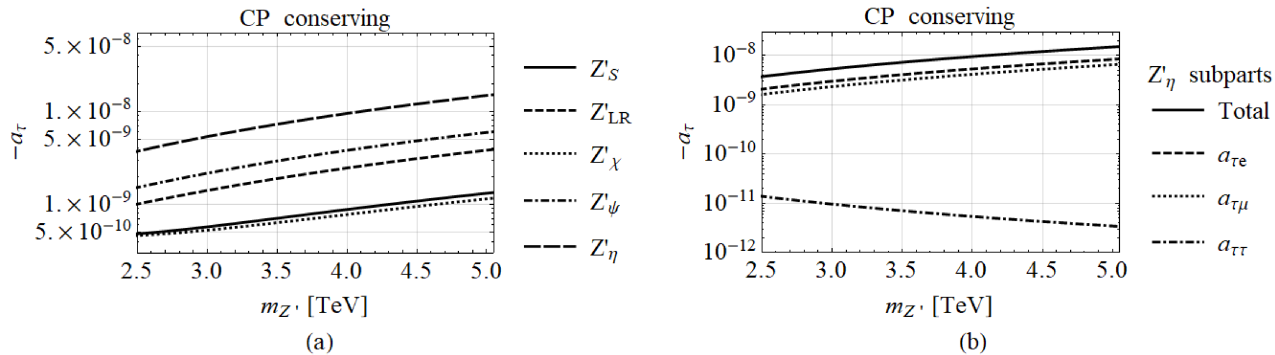}
\caption{\emph{CP} conservation: tau static anomalous magnetic dipole moment. (a) Contributions of the $Z'$ gauge bosons. (b) Main contribution due to $Z'_\eta$ and its subparts.}
\label{FIGURE-CPC-tau.eps}
\end{figure}
The $a_\tau$ results in the \emph{CP}-c scenario as a function of the $Z^\prime$ gauge boson mass, for the interval $m_{Z'}=[2.5,5]$ TeV, are illustrated in Fig.~\ref{FIGURE-CPC-tau.eps}. In Fig.~\ref{FIGURE-CPC-tau.eps}(a) the contributions from the various $Z'$ gauge bosons are shown; the highest signal is provided by the $Z'_\eta$ boson, which goes from $10^{-9}$ to $10^{-8}$, barely one order of magnitud below the SM electroweak (EW) contribution $a_\tau(\text{EW})=5.52\times10^{-7}$ with opposite sign, while the lowest one corresponds to the $Z'_\chi$ boson, which ranges between $10^{-10}$ and $10^{-9}$. In Fig.~\ref{FIGURE-CPC-tau.eps}(b) the main contribution belonging to $Z'_\eta$ is detailed, where the $a_{\tau e}$ and $a_{\tau\mu}$ components essentially represent the signal, while $a_{\tau\tau}$ is three orders of magnitude below. To contextualize our results, we cite some predictions of $a_\tau$ in some extended models. The estimations for $a_\tau$ coming from two-Higgs doublet models (THDMs)~\cite{atauTHDMs}, the minimal supersymmetric Standard Model (MSSM)~\cite{atauMSSM}, and the unparticle model (UM)~\cite{atauUM} are of the order of $10^{-6}$, whereas for leptoquark models, $a_\tau$ can be as high as $10^{-8}$~\cite{atauLQM}, which coincides with the strongest prediction of the simplest little Higgs model~\cite{atauSLHM}.

\subsubsection{CP violation: $a_\tau$ and $d_\tau$}
\begin{figure}[htb!]
\centering
\includegraphics[width=17cm]{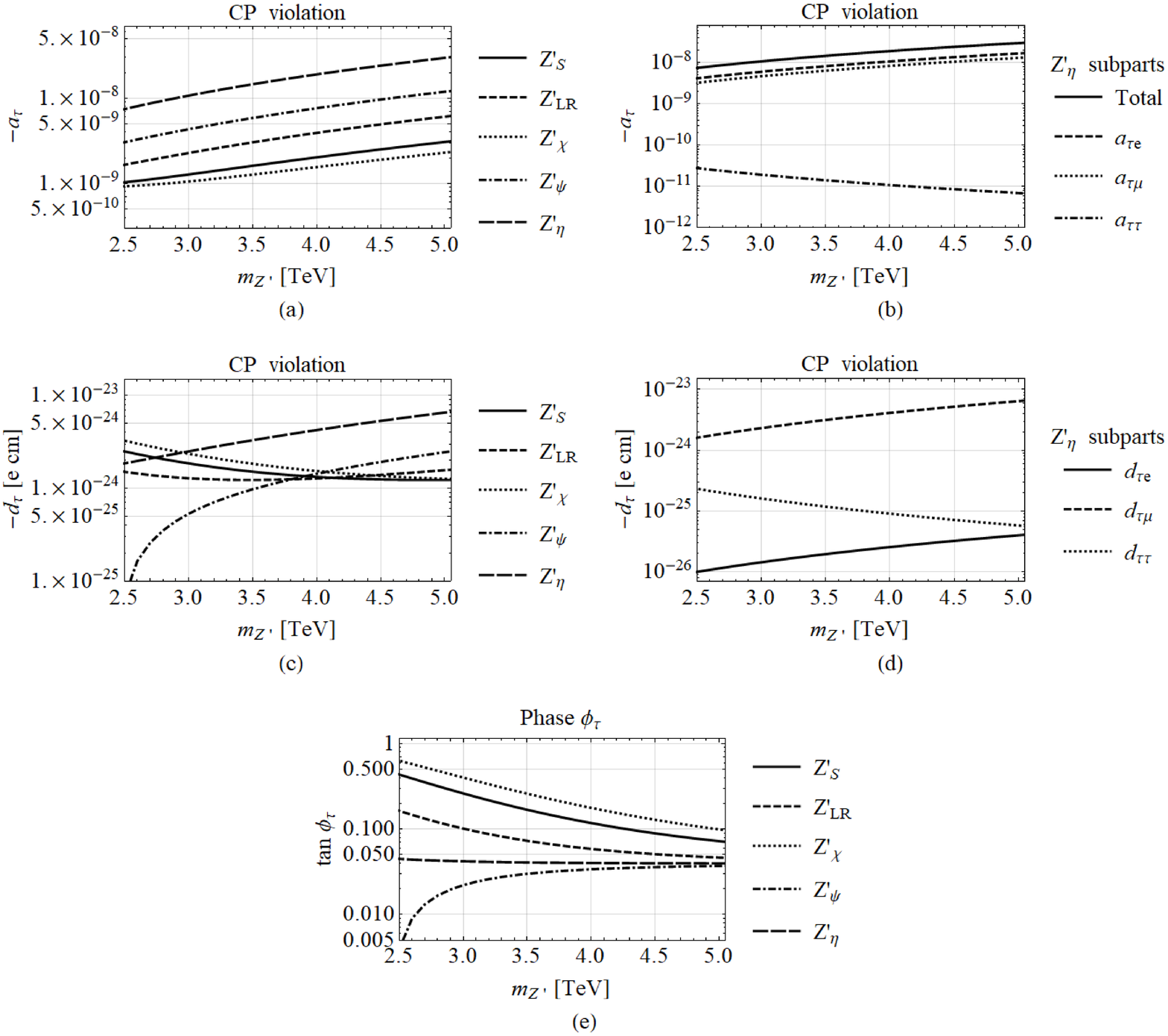}
\caption{\emph{CP} violation: tau static electromagnetic dipole moments. (a) $a_\tau$, contributions of the $Z'$ gauge bosons. (b) $a_\tau$, main contribution due to $Z'_\eta$ and its subparts; (c) $d_\tau$, contributions of the $Z'$ gauge bosons. (d) $d_\tau$, main contribution due to $Z'_\eta$ and its subparts. (e) Phase between EDM and MDM.}
\label{FIGURE-CPV-tau.eps}
\end{figure}

For the \emph{CP}-v analysis, we follow the scenario: $\text{Re}\Omega_L\neq 0, \ \text{Im}\Omega_L= 0,\ \text{Re}\Omega_R=0,\ \text{Im}\Omega_R\neq 0$. Figure~\ref{FIGURE-CPV-tau.eps} presents the results of the tau MDM and EDM in the \emph{CP}-v case. The MDM ($a_\tau$) is displayed in Figs.~\ref{FIGURE-CPV-tau.eps}(a) and (b): in (a), the contributions from the different $Z'$ gauge bosons essentially reproduce the same signals as in the \emph{CP}-c case but are slightly enhanced, and also the $Z'_\eta$ prediction is the leading signal, being of the order of $10^{-8}$, and the $Z'_\chi$ signal is the minor one reaching $10^{-9}$; in (b), the components of the main signal ($Z'_\eta$) are displayed.

On the other hand, in Figs.~\ref{FIGURE-CPV-tau.eps}(c) and \ref{FIGURE-CPV-tau.eps}(d), the EDM of the tau lepton is displayed. In (c), the strongest prediction corresponds to the $Z^\prime_\eta$ gauge boson, while the lower is offered by $Z^\prime_\psi$ ($Z^\prime_S$) in the interval $m_{Z^\prime}=[2_\cdot5,3_\cdot9)$ TeV ($m_{Z^\prime}=[3_\cdot9,5]$ TeV), respectively; in (d), the subparts of the main prediction are shown, where $d_{\tau\mu}$ represents the main contribution.

In the Fig.~\ref{FIGURE-CPV-tau.eps}(e) the $\phi_\tau$ phase (see Eq.~(\ref{chiral-phase})) is depicted and represents the relative size of the EDM respect to the MDM. From this plot we can appreciate that the $Z'_\chi$ signal provides the values closest to 1, while the smallest one corresponds to $Z'_\psi$.

\subsection{Chromoelectromagnetic dipole moments}
\label{Sec:chromo}

The effective Lagrangian that comprises chromoelectromagnetic dipole moments for quarks, $f=q$, is
\begin{align}\label{lagrangian-quarks}
\mathcal{L}_\text{eff}=&\ -\frac{1}{2}T^a\bar{f}\sigma^{\mu\nu}(\mu+id\gamma^5)fG_{\mu\nu}^a,
\end{align}
where $T^a$ is the color generator, $\mu$ is the chromomagnetic form factor and $d$ the chromoelectric form factor, and $G_{\mu\nu}^a$ is the gluon strength field. The CMDM $\mu_f$ and the CEDM $d_f$ \cite{PDG,Bounds-CMDM-CEDM,Bernreuther:2013aga} can be defined dimensionless as $\hat{\mu}_f$ and $\hat{d}_f$:
\begin{equation}\label{chromoelectromagnetic-quark}
\mu\equiv\frac{g_s}{m_f}\hat{\mu}_f \quad , \quad d\equiv \frac{g_s}{m_f}\hat{d}_f \ .
\end{equation}
In analogy to the electromagnetic dipoles given in (\ref{electromagnetic-lepton}), then, $\mu\equiv F_M$ and $d\equiv F_E$.

In general, the chromoelectromagnetic dipoles are complex quantities. The current available experimental bounds from PDG~\cite{PDG,Bounds-CMDM-CEDM} to the quark top dipoles are $-0.053<\text{Re}~\hat{\mu}_t<0.026$ and $-0.068<\text{Im}~\hat{d}_t<0.067$, obtained in the context of an off-shell gluon-top vertex with a timelike scenario $q^2>0$ in hadronic $t\bar{t}$ production, where absorptive imaginary parts for both dipoles are expected. On the other hand, in contrast to the fermion electromagnetic dipole moments defined with the on-shell photon, $q^2=0$, in perturbative QCD, the chromoelectric dipoles cannot be defined on shell because this does not make sense, they are not quantities physically sensitive to that case, and instead, they must be measured off shell at large gluon momentum transfer $q^2\neq 0$ \cite{Choudhury:2014lna}.

To properly compare our obtained results in this section with the SM predictions, we have to revisit the chromomagnetic dipole moment of the top quark in the SM at the one-loop level, for which we have chosen to evaluate at $q^2=\pm m_Z^2$. We must keep in mind that the weak-mixing angle, $\sin^2\theta_W(m_Z)=0.23122$, and alpha strong, $\alpha_s(m_Z)=0.1181$, are experimentally known at the scale of the $Z$ mass \cite{PDG}. Reference \cite{Choudhury:2014lna} only calculated the $q^2=-m_Z^2$ case, and the authors allowed a small mass of the virtual gluons; nevertheless, we cannot reproduce their Eq.~(9). On the other hand, we agree with these authors in the observation that the three-gluon vertex diagram considered in Ref.~\cite{Martinez:2007qf} was not properly calculated; such a diagram is in fact divergent when $q^2=0$. In advance, our derived results given in Table \ref{TABLE-chromo-top-SM} show that the contributions at $q^2=\pm m_Z^2$ coming from the virtual particles $\gamma$, $Z$, $H$, and $g$ barely change, while the $W$ contribution changes sign for its real part; besides, the three-gluon vertex contribution, at which we refer as $3g$, cures its ill behavior when it is off shell. Furthermore, we have found that the contributions from $W$ and $3g$ provide imaginary parts, and as far as we know, this characteristic has not been carefully reported in the literature.
Notice that the on-shell gluon scenario, $q^2=0$, for $\gamma$, $Z$, $H$, and $g$, whose diagrams have in common the same quark as virtual and off shell, serves as an approximate or rough average with respect to the $q^2=\pm m_Z^2$ evaluations.
These results will soon be presented in depth elsewhere, where in addition we will show that in our calculations it is unnecessary to consider a small mass of the virtual gluons~\cite{Us}.

\begin{table}
  \centering
\begin{tabular}{|c|c|c|c|}\hline
\multirow{2}{*}{$\hat{\mu}_t$} & \multicolumn{3}{c|}{$q^2$} \\
  \cline{2-4}
& $-m_Z^2$   & 0 & $m_Z^2$  \\
\hline
$\gamma$        & $2.47\times10^{-4} $ & $2.58\times10^{-4}$   & $2.71\times10^{-4}$    \\
$Z$             & $-1.79\times10^{-3}$ & $-1.85\times10^{-3}$  & $-1.91\times10^{-3}$   \\
$W$             & $-3.42\times10^{-5}-9.43\times10^{-4} i$ & $-2.64\times10^{-6}-1.23\times10^{-3}i$  & $1.44\times10^{-4}-1.19\times10^{-3} i$  \\
$H$             & $1.89\times10^{-3}$  & $1.95\times10^{-3}$   & $2.02\times10^{-3}$  \\
$g$             & $-1.50\times10^{-3}$ & $-1.57\times10^{-3}$  & $-1.64\times10^{-3}$  \\
$3g$            & $-2.13\times10^{-2}$ & indeterminate  & $-1.22\times10^{-2}-2.56\times10^{-2} i$ \\
Total           & $-2.24\times10^{-2}-9.43\times10^{-4}i$ & $-1.20\times10^{-3}-1.23\times10^{-3} i$ & $-1.34\times10^{-2}-2.68\times10^{-2}i$ \\
\hline
\end{tabular}
\caption{Anomalous chromomagnetic dipole moment of the top quark at one-loop level in the SM as function of the gluon momentum transfer $q^2=-m_Z^2, 0, m_Z^2$. The total value for $q^2=0$ does not take into account the triple gluon contribution because it diverges.}\label{TABLE-chromo-top-SM}
\end{table}

\subsection{Predictions on the chromoelectromagnetic dipole moments of the top quark induced by FCNCs}

To calculate the chromoelectromagnetic dipoles of the top quark, we are going to consider the gluon off shell with a 4-momentum transfer $q^2=\pm m_Z^2$; nevertheless, despite being aware that the chromodipoles must be computed with $q^2\neq 0$, for comparison purposes, we also are going to evaluate the on-shell scenario ($q^2=0$).
In advance, as it will be shown below, the Re$\hat{\mu}_t(Z')$ and Re$\hat{d}_t(Z')$ are essentially invariant to any of the three cases $q^2=0,\pm m_Z^2$, while only the timelike scenario, $q^2=m_Z^2$, gives rise to Im$\hat{\mu}_t(Z')$ and Im$\hat{d}_t(Z')$.

The chromoelectromagnetic one-loop diagram is analogous to the photon case, as already commented in Sec.~\ref{SECTION-loop-diagram}, except that for the gluon in the loop integral (see Eq.~(\ref{loop})) $eQ_{f_j}$ must be replaced by $g_sT^a$.

The top-quark CMDM is conformed by the contributions
\begin{equation}\label{chromomagnetic-top}
\hat{\mu}_t=\hat{\mu}_{tu}+\hat{\mu}_{tc}+\hat{\mu}_{tt}~,
\end{equation}
and similarly for the top CEDM,
\begin{equation}\label{chromoelectric-top}
\hat{d}_t=\hat{d}_{tc}+\hat{d}_{tc}+\hat{d}_{tt}~,
\end{equation}
where the components are defined in (\ref{chromoelectromagnetic-quark}). Below we are going to present the CEDM in units of $e$cm.

As already commented on above, the Re$\hat{\mu}_t(Z')$ and Re$\hat{d}_t(Z')$ parts are essentially invariant to the $q^2=0,\pm m_Z^2$ scenarios, and the differences are away from the significant numbers; hence the same form factors $F_M$ and $F_E$ derived for the on-shell case in Eqs.~(\ref{magnetic}) and (\ref{electric}), respectively, allow us now to compute Re$\hat{\mu}_t(Z')$=$F_M$ and Re$\hat{d}_t(Z')$=$F_E$. These form factors were already used to evaluate the tau static dipoles, where $m_\tau\ll m_{Z'}$, but they are still appropriate to evaluate the top-quark dipoles because $m_t\ll m_{Z'}$; we have crossed-checked this by comparing with the unapproximated form factors, and they match excellently. On the other side, the imaginary parts of the chromoelectromagnetic top-quark dipoles, that arise when $q^2=m_Z^2$, are computed with the exact form factors.

\subsubsection{CP conservation: $\hat{\mu}$}
\begin{center}
\begin{figure}[htb!]
\includegraphics[width=17cm]{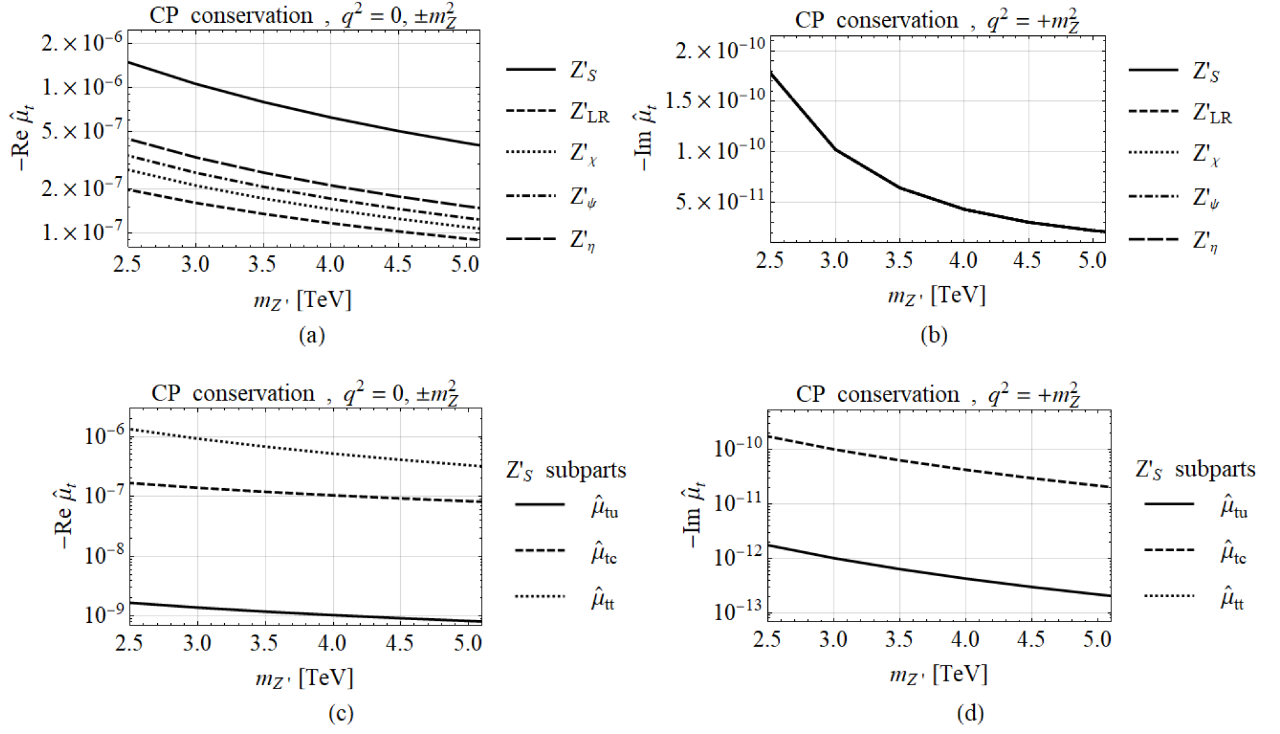}
\caption{\emph{CP} conservation: top magnetic dipole moment.
(a) Contributions of the $Z'$ gauge bosons from different models to the
Re$\hat{\mu}_t$ generated by $q^2=0,\pm m_Z^2$ and (b) Im$\hat{\mu}_t$ generated by $q^2=m_Z^2$, where all the different $Z'$ bosons share essentially the same imaginary value.
(c) Main contribution due to $Z'_S$ to Re$\hat{\mu}_t$ and (d) Im$\hat{\mu}_t$ which arise only from the nondiagonal subparts $\hat{\mu}_{tu,tc}$.}
\label{FIGURE-CPC-top}
\end{figure}
\end{center}

\begin{center}
\begin{figure}[htb!]
\includegraphics[width=16cm]{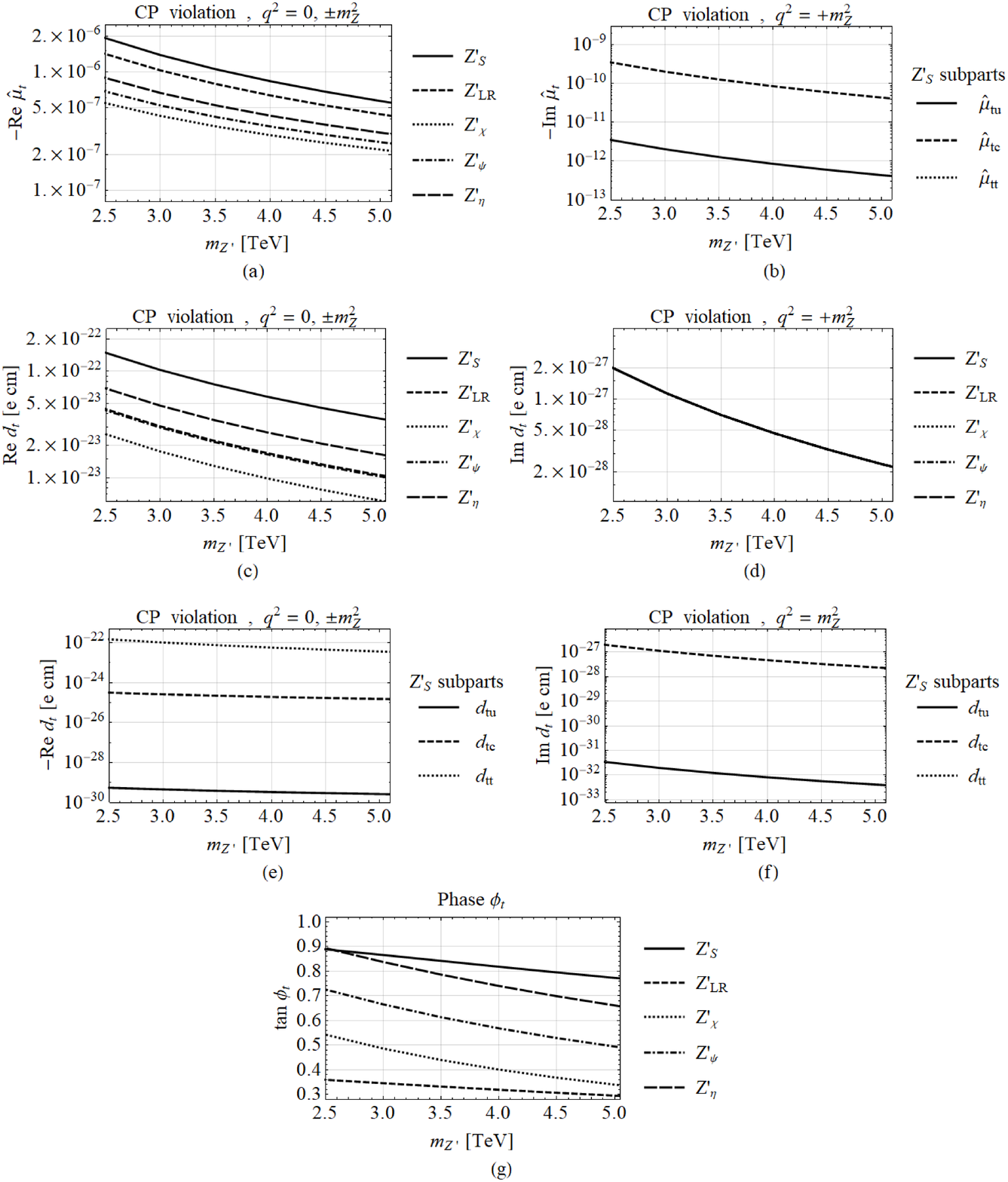}
\caption{\emph{CP} violation: top electromagnetic dipole moments. (a) Contributions of the $Z'$ gauge bosons from different models to Re$\hat{\mu}_t$ and (b) the Im$\hat{\mu}_t$. (c) The Re$\hat{d}_t$ coming from the different $Z'$ and (d) the Im$\hat{d}_t$. In (e) and (f) the respective real and imaginary parts generated by the subparts of the main contributor $Z'_S$. (g) The phase.
}
\label{FIGURE-CPV-top}
\end{figure}
\end{center}

For the analysis of the \emph{CP}-c we follow the scenario: $\text{Re}\Omega_L\neq 0, \ \text{Im}\Omega_L\neq 0,\ \text{Re}\Omega_R=0,\ \text{Im}\Omega_R=0$. Since the coupling parameters $\Omega_{L,R~tc}, \Omega_{L,R~tu}$ in $\hat{\mu}_t$ were estimated in Ref.~\cite{NCS1}, we follow that procedure updated to the current permitted values for the $Z^\prime$ mass, where Eqs.~(\ref{ecD0}) are employed. In Figs.~\ref{FIGURE-CPC-top}~(a)-(d), the results for the $\hat{\mu}_t$ in the \emph{CP}-c case are shown as a function of the $Z^\prime$ boson mass, $m_{Z'}=[2.5,5]$ TeV: in (a) the contributions to Re$\hat{\mu}_t$ from the different $Z'$ gauge bosons are presented, where the leading contribution is due to the $Z'_S$ gauge boson, which decreases from $10^{-6}$ to $10^{-7}$ in the interval, while $Z'_{LR}$ is responsible for the smallest values, which go from $10^{-7}$ to $10^{-8}$; in (b) the Im$\hat{\mu}_t$ is shown, where all the different $Z'$ bosons share the same imaginary value; in (c) the subparts of the main contributor, $Z'_S$, with its Re$\hat{\mu}_t$ are displayed, being  $\hat{\mu}_{tt}$ the highest one, while $\hat{\mu}_{tc}$ is three orders of magnitude below; in (d) the subparts of $Z'_S$ that contribute to Im$\hat{\mu}_t$ are exhibited, which are generated only by the nondiagonals $\hat{\mu}_{tu}$ and $\hat{\mu}_{tc}$. Now, we can compare with the closest SM value, which corresponds to $\hat{\mu}_t(W)=-3.419\times10^{-5}$$-9.434\times10^{-4}i$, when $q^2=-m_Z^2$, where the real part of the $Z'_S$ starts one order of magnitude below, while the imaginary part is six orders lower.

\subsubsection{CP violation: $\hat{\mu}$ and $\hat{d}$}

The \emph{CP}-v analysis is carried out according to the scenario: $\text{Re}\Omega_L\neq 0, \ \text{Im}\Omega_L= 0,\ \text{Re}\Omega_R=0,\ \text{Im}\Omega_R\neq 0$. The $\hat{\mu}_t$ results are presented in Figs.~\ref{FIGURE-CPV-top}(a)-(b): in (a), the contributions from the different $Z'$ gauge bosons can be appreciated, where the $Z'_S$ provides again the highest signal to Re$\hat{\mu}_t$, but a little higher than in the \emph{CP}-c case, being $10^{-6}$ in $m_{Z'}=[2.5,3.5)$ and $10^{-7}$ in $m_{Z'}=[3.5,5]$ TeV. Here, $Z'_\chi$ produces the lowest value, while in the \emph{CP}-c scenario was due to the $Z^\prime_{LR}$; in (b) the imaginary part remains in the order of $10^{-10}$. The corresponding subparts due to the main contributor, $Z'_S$, behave in a way similarly as in the \emph{CP}-c case; we do not show them. Once again, these values are just below the SM subpart coming from the $W$ gauge boson diagram.

Now, we turn our attention to the CEDM, which does not exist in the SM at the one-loop level. The results for $\hat{d}_t$ are shown in Figs.~\ref{FIGURE-CPV-top}(c)-(f) in units of $e$cm. Figure (c) displays the contributions to Re$\hat{d}_t$ from the different $Z'$ gauge bosons, and again the same results are provided by the scenarios $q^2=0,\pm m_Z^2$, the differences are away from the significant numbers, and also the $Z^\prime_S$ is responsible for the highest signal, being $10^{-22}$ $e\,$cm in $m_{Z'}=[2.5,3.2)$ and $10^{-23}$ $e\,$cm in $m_{Z'}=[3.2,5]$ TeV. In contrast, the $Z'_\chi$ boson offers the lowest signal which is one order of magnitude below the $Z'_S$ one; in (d) the corresponding imaginary part is exhibited; in (e) we can see that the diagonal Re$\hat{\mu}_{tt}$ is the responsible for the highest value; in (f), the nondiagonal subparts generate the imaginary part. Finally, the $\phi_t$ phase is presented in Fig.~\ref{FIGURE-CPV-top}(g), where the $Z'_S$ boson yields the most intense \emph{CP}-violation behavior, whereas the lesser one is due to the $Z'_{LR}$ boson.

\section{Conclusions}
The new physics effects due to the possible presence of FCNCs mediated by a new neutral massive gauge boson, identified as $Z^\prime$, have been studied on the MDM (EDM) of the tau lepton and the CMDM (CEDM) of the top quark. The theoretical framework corresponds to the most general renormalizable Lagrangian that includes flavor violation mediated by a gauge boson type $Z^\prime$, which can be induced in grand unification models. By using constraints, calculated in a previous work, of the lepton flavor-violating couplings $Z^\prime\tau\mu$ and $Z^\prime\tau e$, coming from experimental bounds for the lepton flavor-violating $\tau\to\mu\mu^+\mu^-$ and  $\tau\to\mu e^+e^-$ decays, the MDM ($a_\tau$) and the EDM ($d_\tau$) of the tau lepton were estimated. Specifically, for the \emph{CP} conservation case, where only $a_\tau$ is induced, we found that $|a_\tau|\sim10^{-8}$ at best for the $Z^\prime_\eta$ boson, which is of the same order of magnitude as the respective predictions in the leptoquark models and the simplest little Higgs model; the remanning $Z^\prime$ bosons offer values for $|a_\tau|$ between $10^{-10}$ and $10^{-9}$. Besides, for the \emph{CP} violation case, also $|a_\tau|$ can be as high as $10^{-8}$ for the $Z^\prime_\eta$ boson, while the other $Z^\prime$ boson contributions can reach $10^{-9}$; in relation to the EDM ($d_\tau$), the highest prediction for the $|d_\tau|$ corresponds to the $Z^\prime_\eta$, with $|d_\tau|$ being of the order of $10^{-24}$ $e\,$cm, whereas the SM prediction is less than $10^{-34}$ $e\,$cm.

In addition, by considering the results of a previous work in which the strength of the $Z^\prime tc$ and $Z^\prime tu$ couplings were estimated through the $D^0-\bar{D^0}$ mixing system, the FCNC predictions for the CMDM ($\hat{\mu}_t$) and the CEDM ($\hat{d}_t$) of the top quark were calculated. We have revisited the SM predictions in order to be able to compare the results of the chromodipoles induced by FCNCs, for which we have considered the off-shell gluon 4-momentum transfer $q^2=\pm m_Z^2$, where imaginary contributions are generated. For the \emph{CP}-conservation and \emph{CP}-violation scenarios, the main signal is offered by the $Z'_S$ boson, being of the order of $-$Re$\hat{\mu}_t\sim$$10^{-6}-10^{-7}$ and $-$Im$\hat{\mu}_t\sim$ $10^{-10}-10^{-11}$, where the real part value starts barely one order of magnitude below the SM prediction due to the $W$ boson. The CEDM, $\hat{d}_t$, is estimated to be in the interval $-$Re$\hat{d}_t\sim$ $10^{-23}-10^{-22}$ $e\,$cm and $-$Im$\hat{d}_t\leq 10^{-27}$ $e\,$cm, where signals provided by the $Z^\prime_S$ boson correspond to the best situation. All our predictions agree with the current experimental limits.

\section*{Acknowledgments}
This work has been partially supported by SNI-CONACYT and CIC-UMSNH. J. M. thanks to the Catedras CONACYT Program support.


\begin{thebibliography}{99}
\bibitem{SMqsectorsup} For instance, see G. Eilam, J. L. Hewett, and A. Soni, \href{https://doi.org/10.1103/PhysRevD.44.1473}{{\textcolor{blue}{Phys. Rev. D \textbf{44}, 1473 (1991)}}}; \href{https://doi.org/10.1103/PhysRevD.59.039901}{{\textcolor{blue}{\textbf{59}, 039901(E) (1998)}}}; N. G. Deshpande, B. Margolis, and H. D. Trottier, \href{https://doi.org/10.1103/PhysRevD.45.178}{{\textcolor{blue}{Phys. Rev. D \textbf{45}, 178 (1992)}}}; B. Mele, S. Petrarca, and A Soddu, \href{https://doi.org/10.1016/S0370-2693(98)00822-3}{{\textcolor{blue}{Phys. Lett. B \textbf{435}, 401 (1998)}}}; A. Cordero-Cid, J. M. Hern\'andez, G. Tavares-Velasco, and J. J. Toscano, \href{https://doi.org/10.1103/PhysRevD.73.094005}{{\textcolor{blue}{Phys. Rev. D \textbf{73}, 094005 (2006)}}}; G. Eilam, M. Frank, and I. Turan, \href{https://doi.org/10.1103/PhysRevD.73.053011}{{\textcolor{blue}{Phys. Rev. D \textbf{73}, 053011 (2006)}}}; \href{https://doi.org/10.1103/PhysRevD.74.035012}{{\textcolor{blue}{\textbf{74}, 035012 (2006)}}}.

\bibitem{EYS} J.~I.~Aranda, F.~Ram\'irez-Zavaleta, J.~J.~Toscano, and E.~S.~Tututi, \href{https://doi.org/10.1103/PhysRevD.78.017302}{{\textcolor{blue}{Phys.\ Rev.\ D {\bf 78}, 017302 (2008)}}}; J.~I.~Aranda, A.~Flores-Tlalpa, F.~Ram\'irez-Zavaleta, F.~J.~Tlachino, J.~J.~Toscano, and E.~S.~Tututi, \href{https://doi.org/10.1103/PhysRevD.79.093009}{{\textcolor{blue}{Phys.\ Rev.\ D {\bf 79}, 093009 (2009)}}}; J.~I.~Aranda, A.~Cordero-Cid, F.~Ram\'irez-Zavaleta, J.~J.~Toscano, and E.~S.~Tututi, \href{https://doi.org/10.1103/PhysRevD.81.077701}{{\textcolor{blue}{Phys.\ Rev.\ D {\bf 81}, 077701 (2010)}}}; A.~Fern\'andez, C.~Pagliarone, F.~Ram\'irez-Zavaleta, and J.~J.~Toscano, \href{https://doi.org/10.1088/0954-3899/37/8/085007}{{\textcolor{blue}{J.\ Phys.\ G {\bf 37}, 085007 (2010)}}}; J.~I.~Aranda, J.~Monta\~no, F.~Ram\'irez-Zavaleta, J.~J.~Toscano, and E.~S.~Tututi, \href{https://doi.org/10.1103/PhysRevD.82.054002}{{\textcolor{blue}{Phys.\ Rev.\ D {\bf 82}, 054002 (2010)}}}; A. Das, T. Nomura, H. Okada, and S. Roy, \href{https://doi.org/10.1103/PhysRevD.96.075001}{{\textcolor{blue}{Phys. Rev. D \textbf{96}, 075001 (2017)}}}.

\bibitem{arhrib} A. Arhrib \emph{et al.}, \href{https://doi.org/10.1103/PhysRevD.73.075015}{{\textcolor{blue}{Phys. Rev. D \textbf{73}, 075015 (2006)}}}.

\bibitem{NCS1} J.~I.~Aranda, F.~Ram\'irez-Zavaleta, J.~J.~Toscano, and E.~S.~Tututi, \href{https://doi.org/10.1088/0954-3899/38/4/045006}{{\textcolor{blue}{J.\ Phys.\ G {\bf 38}, 045006 (2011)}}}.

\bibitem{NCS2} J.~I.~Aranda, J.~Monta\~no, F.~Ram\'irez-Zavaleta, J.~J.~Toscano, and E.~S.~Tututi, \href{https://doi.org/10.1103/PhysRevD.86.035008}{{\textcolor{blue}{Phys.\ Rev.\ D {\bf 86}, 035008 (2012)}}}.

\bibitem{neutrinos} R. Becker-Szendy \emph{et al}., \href{https://doi.org/10.1016/0920-5632(94)00765-N}{{\textcolor{blue}{Nucl. Phys. Proc. Suppl. \textbf{38}, 331 (1995)}}}; Y. Fukuda \emph{et al}., \href{https://doi.org/10.1016/0370-2693(94)91420-6}{{\textcolor{blue}{Phys. Lett. B \textbf{335}, 237 (1994)}}}; \href{https://doi.org/10.1103/PhysRevLett.81.1562}{{\textcolor{blue}{Phys. Rev. Lett. \textbf{81}, 1562 (1998)}}}; H. Sobel, \href{https://doi.org/10.1016/S0920-5632(00)00932-4}{{\textcolor{blue}{Nucl. Phys. Proc. Suppl. \textbf{91}, 127 (2001)}}}; M. Ambrossio \emph{et al}., \href{https://doi.org/10.1016/S0370-2693(03)00806-2}{{\textcolor{blue}{Phys. Lett. B \textbf{566}, 35 (2003)}}}; M. Apollonio \emph{et al}., \href{https://doi.org/10.1140/epjc/s2002-01127-9}{{\textcolor{blue}{Eur. Phys. J. C \textbf{27}, 331 (2003)}}}; M. B. Smy \emph{et al}., \href{https://doi.org/10.1103/PhysRevD.69.011104}{{\textcolor{blue}{Phys. Rev. D \textbf{69}, 011104(R) (2004)}}}; S. N. Ahmed \emph{et al}., \href{https://doi.org/10.1103/PhysRevLett.92.181301}{{\textcolor{blue}{Phys. Rev. Lett. \textbf{92}, 181301 (2004)}}}; Y. Ashie \emph{et al}., \href{https://doi.org/10.1103/PhysRevLett.93.101801}{{\textcolor{blue}{Phys. Rev. Lett. \textbf{93}, 101801 (2004)}}}; E. Aliu \emph{et al}., \href{https://doi.org/10.1103/PhysRevLett.94.081802}{{\textcolor{blue}{Phys. Rev. Lett. \textbf{94}, 081802 (2005)}}}; Y. Ashie \emph{et al}., \href{https://doi.org/10.1103/PhysRevD.71.112005}{{\textcolor{blue}{Phys. Rev. D \textbf{71}, 112005 (2005)}}}; W. W. M. Allison \emph{et al}., \href{https://doi.org/10.1103/PhysRevD.72.052005}{{\textcolor{blue}{Phys. Rev. D \textbf{72}, 052005 (2005)}}}; P. Adamson \emph{et al}., \href{https://doi.org/10.1103/PhysRevD.73.072002}{{\textcolor{blue}{Phys. Rev. D \textbf{73}, 072002 (2006)}}}.

\bibitem{langacker1} M. Cveti\v{c}, P. Langacker, and B. Kayser, \href{https://doi.org/10.1103/PhysRevLett.68.2871}{{\textcolor{blue}{Phys. Rev. Lett. \textbf{68}, 2871 (1992)}}}; M. Cveti\v{c} and P. Langacker, \href{https://doi.org/10.1103/PhysRevD.54.3570}{{\textcolor{blue}{Phys. Rev. D \textbf{54}, 3570 (1996)}}}; M. Cveti\v{c}  \emph{et al.}, \href{https://doi.org/10.1103/PhysRevD.56.2861}{{\textcolor{blue}{Phys. Rev. D \textbf{56}, 2861 (1997)}}}; \href{https://doi.org/10.1103/PhysRevD.58.119905}{{\textcolor{blue}{\textbf{58}, 119905(E) (1998)}}}; M. Masip and A. Pomarol, \href{https://doi.org/10.1103/PhysRevD.60.096005}{{\textcolor{blue}{Phys. Rev. D \textbf{60}, 096005 (1999)}}}; N. Arkani-Hamed, A. G. Cohen, and H. Georgi, \href{https://doi.org/10.1016/S0370-2693(01)00741-9}{{\textcolor{blue}{Phys. Lett. B \textbf{513}, 232 (2001)}}}; N. Arkani-Hamed, A. G. Cohen, E. Katz, and A. E. Nelson, \href{https://doi.org/10.1088/1126-6708/2002/07/034}{{\textcolor{blue}{JHEP \textbf{07}, 034 (2002)}}}; T. Han, H. E. Logan, B. McElrath, and L.-T. Wang, \href{https://doi.org/10.1103/PhysRevD.67.095004}{{\textcolor{blue}{Phys. Rev. D \textbf{67}, 095004 (2003)}}}; C. T. Hill and E. H. Simmons, \href{https://doi.org/10.1016/S0370-1573(03)00140-6}{{\textcolor{blue}{Phys. Rept. \textbf{381}, 235 (2003)}}}; \href{https://doi.org/10.1016/j.physrep.2003.10.002}{{\textcolor{blue}{\textbf{390}, 553 (2004)}}}; J. Kang and P. Langacker, \href{https://doi.org/10.1103/PhysRevD.71.035014}{{\textcolor{blue}{Phys. Rev. D \textbf{71}, 035014 (2005)}}}; B. Fuks \emph{et al.}, \href{https://doi.org/10.1016/j.nuclphysb.2008.01.017}{{\textcolor{blue}{Nucl. Phys. \textbf{B797}, 322 (2008)}}}; J. Erler \emph{et al.}, \href{https://doi.org/10.1088/1126-6708/2009/08/017}{{\textcolor{blue}{JHEP \textbf{08}, 017 (2009)}}}; M. Goodsell \emph{et al.}, \href{https://doi.org/10.1088/1126-6708/2009/11/027}{{\textcolor{blue}{JHEP \textbf{11}, 027 (2009)}}}; P. Langacker, \href{https://doi.org/10.1063/1.3327671}{{\textcolor{blue}{AIP Conf. Proc. \textbf{1200}, 55 (2010)}}}.

\bibitem{robinett} R. W. Robinett and Jonathan L. Rosner, \href{https://doi.org/10.1103/PhysRevD.26.2396}{{\textcolor{blue}{Phys. Rev. D \textbf{26}, 2396 (1982)}}}.

\bibitem{langacker2} P. Langacker and M. Luo, \href{https://doi.org/10.1103/PhysRevD.45.278}{{\textcolor{blue}{Phys. Rev. D \textbf{45}, 278 (1992)}}}.

\bibitem{leike} A. Leike, \href{https://doi.org/10.1016/S0370-1573(98)00133-1}{{\textcolor{blue}{Phys. Rept. \textbf{317}, 143 (1999)}}}.

\bibitem{perez-soriano} M. A. P\'erez and M. A. Soriano, \href{https://doi.org/10.1103/PhysRevD.46.284}{{\textcolor{blue}{Phys. Rev. D \textbf{46}, 284 (1992)}}}.

\bibitem{ATLAS1} M. Aaboud \emph{et al.} [The ATLAS collaboration], \href{https://doi.org/10.1016/j.physletb.2016.08.055}{{\textcolor{blue}{Phys. Lett. B \textbf{761}, 372 (2016)}}}.

\bibitem{CMS1} V. Khachatryan \emph{et al.} [CMS Collaboration], \href{https://doi.org/10.1007/JHEP04(2015)025}{{\textcolor{blue}{JHEP \textbf{04}, 025 (2015)}}}.

\bibitem{Scalars-ATLAS-CMS} M. Aaboud \emph{et al.} [The ATLAS collaboration], \href{https://doi.org/10.1007/JHEP09(2016)001}{{\textcolor{blue}{JHEP \textbf{09}, 001 (2016)}}}; V. Khachatryan \emph{et al.} [CMS Collaboration], \href{https://doi.org/10.1103/PhysRevLett.117.051802}{{\textcolor{blue}{Phys. Rev. Lett. \textbf{117}, 051802 (2016)}}}.

\bibitem{PDG}
  M. Tanabashi \emph{et al.} [Particle Data Group], \href{https://doi.org/10.1103/PhysRevD.98.030001}{{\textcolor{blue}{Phys. Rev. D \textbf{98}, 030001 (2018)}}}.

\bibitem{sahoo1} S. Sahoo, C. K. Das, and L. Maharana, \href{https://doi.org/10.1142/S0217751X11053936}{{\textcolor{blue}{Int. J. Mod. Phys. A \textbf{26}, 3347 (2011)}}}; S. Sahoo, M. Kumar, and D. Banerjee, \href{https://doi.org/10.1142/S0217751X13500607}{{\textcolor{blue}{Int. J. Mod. Phys. A \textbf{28}, 1350060 (2013)}}}.

\bibitem{cabarcas} J. M. Cabarcas, J. Duarte, and J.-Alexis Rodriguez, \href{https://doi.org/10.1142/S0217751X14500158}{{\textcolor{blue}{Int. J. Mod. Phys. A \textbf{29}, 1450015 (2014)}}}.

\bibitem{yue} Chong-Xing Yue and Man-Lin Cui, \href{https://doi.org/10.1016/j.nuclphysb.2014.08.014}{{\textcolor{blue}{Nucl. Phys. \textbf{B887}, 371 (2014)}}}.

\bibitem{langacker-rmp} P. Langacker, \href{https://doi.org/10.1103/RevModPhys.81.1199}{{\textcolor{blue}{Rev. Mod. Phys. \textbf{81}, 1199 (2009)}}}.

\bibitem{amdmtaureview} S. Eidelman and M. Passera, \href{https://doi.org/10.1142/S0217732307022694}{{\textcolor{blue}{Mod. Phys. Lett. A \textbf{22}, 159 (2007)}}}.

\bibitem{g-2-e} G. W. Bennett \emph{et al.} [Muon g-2 Collaboration], \href{https://doi.org/10.1103/PhysRevD.73.072003}{{\textcolor{blue}{Phys. Rev. D \textbf{73}, 072003 (2006)}}}.

\bibitem{g-2-t} T. Blum \emph{et al.}, \href{https://arxiv.org/abs/1311.2198}{{\textcolor{blue}{arXiv:1311.2198[hep-ph]}}}.

\bibitem{g-2-d} M. Procura \emph{et al.}, \href{https://doi.org/10.1051/epjconf/201816600014}{{\textcolor{blue}{EPJ Web of Conferences \textbf{166}, 00014 (2018)}}}.

\bibitem{g-2-ne} J. Grange \emph{et al.} [Muon g-2 Collaboration], \href{https://arxiv.org/abs/1501.06858}{{\textcolor{blue}{arXiv:1501.06858}}};
N. Saito [J-PARC g-2/EDM Collaboration], \href{https://doi.org/10.1063/1.4742078}{{\textcolor{blue}{AIP Conf. Proc. \textbf{1467}, 45 (2012)}}}.

\bibitem{g-2-nt} G. Colangelo, M. Hoferichter, M. Procura, and P. Stoffer, \href{https://doi.org/10.1007/JHEP09(2014)091}{{\textcolor{blue}{JHEP \textbf{09}, 091 (2014)}}}; G. Colangelo, M. Hoferichter, B. Kubis, M. Procura, and P. Stoffer, \href{https://doi.org/10.1016/j.physletb.2014.09.021}{{\textcolor{blue}{Phys. Lett. B \textbf{738}, 6 (2014)}}}.

\bibitem{Belle} K. Inami \emph{et al.} [Belle Collaboration], \href{https://doi.org/10.1016/S0370-2693(02)02984-2}{{\textcolor{blue}{Phys. Lett. B \textbf{551}, 16 (2003)}}}.

\bibitem{GutierrezRodriguez:2006hb}
  A.~Guti\'errez-Rodr\'iguez, M.~A.~Hern\'andez-Ruiz, and M.~A.~P\'erez, \href{https://doi.org/10.1142/S0217751X07036865}{{\textcolor{blue}{Int.\ J.\ Mod.\ Phys.\ A {\bf 22}, 3493 (2007)}}}.

\bibitem{GutierrezRodriguez:2009ns}
  A.~Guti\'errez-Rodr\'iguez, \href{https://doi.org/10.1142/S0217732310032238}{{\textcolor{blue}{Mod.\ Phys.\ Lett.\ A {\bf 25}, 703 (2010)}}}.

\bibitem{Gutierrez-Rodriguez:2013eaa}
  A.~Guti\'errez-Rodr\'iguez, M.~A.~Hern\'andez-Ruiz, and C.~P.~Casta\~neda-Almanza, \href{https://doi.org/10.1088/0954-3899/40/3/035001}{{\textcolor{blue}{J.\ Phys.\ G {\bf 40}, 035001 (2013)}}}.

\bibitem{Bounds-CMDM-CEDM} V. Khachatryan \emph{et al.} [CMS Collaboration], \href{https://doi.org/10.1103/PhysRevD.93.052007}{{\textcolor{blue}{Phys. Rev. D \textbf{93}, 052007 (2016)}}}.

\bibitem{Atwood:1994vm}
  D.~Atwood, A.~Kagan, and T.~G.~Rizzo, \href{https://doi.org/10.1103/PhysRevD.52.6264}{{\textcolor{blue}{Phys.\ Rev.\ D {\bf 52}, 6264 (1995)}}}.

\bibitem{Gaitan:2015aia}
  R.~Gait\'an, E.~A.~Garc\'es, J.~H.~M.~de Oca, and R.~Martinez, \href{https://doi.org/10.1103/PhysRevD.92.094025}{{\textcolor{blue}{Phys.\ Rev.\ D {\bf 92}, 094025 (2015)}}}.

\bibitem{Martinez:2001qs}
  R.~Martinez and J.~A.~Rodriguez, \href{https://doi.org/10.1103/PhysRevD.65.057301}{{\textcolor{blue}{Phys.\ Rev.\ D {\bf 65}, 057301 (2002)}}}.

\bibitem{Aboubrahim:2015zpa}
  A.~Aboubrahim, T.~Ibrahim, P.~Nath, and A.~Zorik, \href{https://doi.org/10.1103/PhysRevD.92.035013}{{\textcolor{blue}{Phys.\ Rev.\ D {\bf 92}, 035013 (2015)}}}.

\bibitem{Martinez:2007qf}
  R.~Mart\'inez, M.~A.~P\'erez, and N.~Poveda, \href{https://doi.org/10.1140/epjc/s10052-007-0457-6}{{\textcolor{blue}{Eur.\ Phys.\ J.\ C {\bf 53}, 221 (2008)}}}.

\bibitem{Appelquist:2004es}
  T.~Appelquist, M.~Piai, and R.~Shrock, \href{https://doi.org/10.1016/j.physletb.2004.06.066}{{\textcolor{blue}{Phys.\ Lett.\ B {\bf 595}, 442 (2004)}}}.

\bibitem{Ibrahim:2011im}
  T.~Ibrahim and P.~Nath, \href{https://doi.org/10.1103/PhysRevD.84.015003}{{\textcolor{blue}{Phys.\ Rev.\ D {\bf 84}, 015003 (2011)}}}.

\bibitem{Hayreter:2013kba}
  A.~Hayreter and G.~Valencia, \href{https://doi.org/10.1103/PhysRevD.88.034033}{{\textcolor{blue}{Phys.\ Rev.\ D {\bf 88}, 034033 (2013)}}}.

\bibitem{Tavares}
  A.~I.~Hern\'andez-Ju\'arez, A.~Moyotl, and G.~Tavares-Velasco, \href{https://doi.org/10.1103/PhysRevD.98.035040}{{\textcolor{blue}{Phys.\ Rev.\ D {\bf 98}, 035040 (2018)}}}.

\bibitem{Choudhury:2014lna}
  I.~D.~Choudhury and A.~Lahiri, \href{https://doi.org/10.1142/S0217732315501138}{{\textcolor{blue}{Mod.\ Phys.\ Lett.\ A {\bf 30}, 1550113 (2015)}}}.

\bibitem{Bermudez:2017bpx}
  R.~Bermudez, L.~Albino, L.~X.~Guti\'errez-Guerrero, M.~E.~Tejeda-Yeomans, and A.~Bashir, \href{https://doi.org/10.1103/PhysRevD.95.034041}{{\textcolor{blue}{Phys.\ Rev.\ D {\bf 95}, 034041 (2017)}}}.

\bibitem{durkin} L. S. Durkin and P. Langacker, \href{https://doi.org/10.1016/0370-2693(86)91594-7}{{\textcolor{blue}{Phys. Lett. B \textbf{166}, 436 (1986)}}}; Y. Y. Komachenko and M. Y. Khlopov, \href{https://inspirehep.net/record/285983?ln=es}{{\textcolor{blue}{Sov. J. Nucl. Phys. \textbf{51}, 692 (1990)}}}; M. Cvetic and P. Langacker, \href{https://doi.org/10.1142/9789814536684}{{\textcolor{blue}{Proceedings of Ottawa 1992: Beyond the standard model 3, 454-458, (1992)}}}; Cheng-Wei Chiang, Yi-Fan Lin, and Jusak Tandean, \href{https://doi.org/10.1007/JHEP11(2011)083}{{\textcolor{blue}{JHEP 11, 083 (2011)}}}.

\bibitem{langacker3} P. Langacker and M. Pl$\mathrm{\ddot{u}}$macher, \href{https://doi.org/10.1103/PhysRevD.62.013006}{{\textcolor{blue}{Phys. Rev. D \textbf{62}, 013006 (2000)}}}; X.-G. He and G. Valencia, \href{https://doi.org/10.1103/PhysRevD.74.013011}{{\textcolor{blue}{Phys. Rev. D \textbf{74}, 013011 (2006)}}}; C.-W. Chiang, N. G. Deshpande, and J. Jiang, \href{https://doi.org/10.1088/1126-6708/2006/08/075}{{\textcolor{blue}{JHEP \textbf{08}, 075 (2006)}}}.

\bibitem{Salam-Mohapatra} J. C. Pati and A. Salam, \href{https://doi.org/10.1103/PhysRevD.10.275}{{\textcolor{blue}{Phys. Rev. D \textbf{10}, 275 (1974)}}}; \href{https://doi.org/10.1103/PhysRevD.11.703.2}{{\textcolor{blue}{\textbf{11}, 703(E) (1975)}}};
R. N. Mohapatra and J. C. Pati,  \href{https://doi.org/10.1103/PhysRevD.11.566}{{\textcolor{blue}{Phys. Rev. D \textbf{11}, 566 (1975)}}}.

\bibitem{Pleitez} F. Pisano and V. Pleitez, \href{https://doi.org/10.1103/PhysRevD.46.410}{{\textcolor{blue}{Phys. Rev. D \textbf{46}, 410 (1992)}}}; P. H. Frampton, \href{https://doi.org/10.1103/PhysRevLett.69.2889}{{\textcolor{blue}{Phys. Rev. Lett. \textbf{69}, 2889 (1992)}}}.

\bibitem{robinett2} Richard W. Robinett and Jonathan L. Rosner, \href{https://doi.org/10.1103/PhysRevD.25.3036}{{\textcolor{blue}{Phys. Rev. D \textbf{25}, 3036 (1982)}}}; \href{https://doi.org/10.1103/PhysRevD.27.679}{{\textcolor{blue}{Phys. Rev. D \textbf{27}, 679(E) (1983)}}}; R. W. Robinett, \href{https://doi.org/10.1103/PhysRevD.26.2388}{{\textcolor{blue}{Phys. Rev. D \textbf{26}, 2388 (1982)}}}.

\bibitem{mar} M. Artuso \textit{et al.}, \href{https://doi.org/10.1103/PhysRevLett.95.251801}{{\textcolor{blue}{Phys. Rev. Lett. \textbf{95}, 251801 (2005)}}}.

\bibitem{Jordy}
    Y.~T.~Chien, V.~Cirigliano, W.~Dekens, J.~de Vries, and E.~Mereghetti, \href{https://doi.org/10.1007/JHEP02(2016)011}{{\textcolor{blue}{JHEP {\bf 02}, 011 (2016)}}}; V. Cirigliano, W. Dekens, J. de Vries, and E. Mereghetti, \href{https://doi.org/10.1103/PhysRevD.94.034031}{{\textcolor{blue}{Phys. Rev. D \textbf{94}, 034031 (2016)}}}.

\bibitem{Roberts:2010zz} B.~L.~Roberts and W.~J.~Marciano, \href{https://doi.org/10.1142/7273}{{\textcolor{blue}{Adv.\ Ser.\ Direct.\ High Energy Phys. {\bf 20}, pp. 1 (2009)}}}.

\bibitem{Jegerlehner:2017gek}
  F.~Jegerlehner, \href{https://doi.org/10.1007/978-3-319-63577-4}{{\textcolor{blue}{Springer Tracts Mod.\ Phys. {\bf 274}, pp.1 (2017)}}}.

\bibitem{atauTHDMs} J. Bernab\'eu, D. Comelli, L. Lavoura, and J. P. Silva, \href{https://doi.org/10.1103/PhysRevD.53.5222}{{\textcolor{blue}{Phys. Rev. D \textbf{53}, 5222 (1996)}}}; D. G\'omez Dumm and G. A. Gonz\'alez-Sprinberg, \href{https://doi.org/10.1007/s100529900185}{{\textcolor{blue}{Eur. Phys. J. C \textbf{11}, 293 (1999)}}}.

\bibitem{atauMSSM} W. Hollik, J. I. Illana, S. Rigolin, C. Schappacher, and D. Stockinger, \href{https://doi.org/10.1016/S0550-3213(99)00201-1}{{\textcolor{blue}{Nucl. Phys. \textbf{B551}, 3 (1999)}}}; W. Hollik, J. I. Illana, C. Schappacher, and D. Stockinger, \href{https://doi.org/10.1016/S0550-3213(99)00396-X}{{\textcolor{blue}{Nucl. Phys. \textbf{B557}, 407 (1999)}}}.

\bibitem{atauUM} A. Moyotl and G. Tavares-Velasco, \href{https://doi.org/10.1103/PhysRevD.86.013014}{{\textcolor{blue}{Phys. Rev. D \textbf{86}, 013014 (2012)}}}.

\bibitem{atauLQM} A. Bola\~nos, A. Moyotl, and G. Tavares-Velasco, \href{https://doi.org/10.1103/PhysRevD.89.055025}{{\textcolor{blue}{Phys. Rev. D \textbf{89}, 055025 (2014)}}}.

\bibitem{atauSLHM} M. A. Arroyo-Ure\~na, G. Hern\'andez-Tom\'e, and G. Tavares-Velasco, \href{https://doi.org/10.1140/epjc/s10052-017-4803-z}{{\textcolor{blue}{Eur. Phys. J. C \textbf{77}, 227 (2017)}}}.

\bibitem{Bernreuther:2013aga}
  W.~Bernreuther and Z.-G.~Si, \href{https://doi.org/10.1016/j.physletb.2013.06.051}{{\textcolor{blue}{Phys.\ Lett.\ B {\bf 725}, 115 (2013)}}}; W.~Bernreuther and Z.-G.~Si, \href{https://doi.org/10.1016/j.physletb.2015.03.035}{{\textcolor{blue}{Phys.\ Lett.\ B {\bf 744}, 413 (2015)}}}.

\bibitem{Us} J. I. Aranda, J. Monta\~no, B. Quezadas-Vivian, F. Ram\'irez-Zavaleta, and E. S. Tututi, paper in preparation.


\end{thebibliography}
\end{document}